\documentstyle[twoside,fleqn,espcrc2]{article}

\newcommand{\nc}{\newcommand}
\nc{\be}{\begin{equation}}
\nc{\ee}{\end{equation}}
\nc{\bea}{\begin{eqnarray}}
\nc{\eea}{\end{eqnarray}}
\nc{\eqn}[1]{{(\ref{#1})}}
\nc{\cA}{{\cal A}}
\nc{\cB}{{\cal B}}
\nc{\cC}{{\cal C}}
\nc{\cD}{{\cal D}}
\nc{\cE}{{\cal E}}
\nc{\cF}{{\cal F}}
\nc{\cG}{{\cal G}}
\nc{\cH}{{\cal H}}
\nc{\cI}{{\cal I}}
\nc{\cJ}{{\cal J}}
\nc{\cK}{{\cal K}}
\nc{\cL}{{\cal L}}
\nc{\cM}{{\cal M}}
\nc{\cN}{{\cal N}}
\nc{\cO}{{\cal O}}
\nc{\cP}{{\cal P}}
\nc{\cQ}{{\cal Q}}
\nc{\cR}{{\cal R}}
\nc{\cS}{{\cal S}}
\nc{\cT}{{\cal T}}
\nc{\cU}{{\cal U}}
\nc{\cV}{{\cal V}}
\nc{\cW}{{\cal W}}
\nc{\cX}{{\cal X}}
\nc{\cY}{{\cal Y}}
\nc{\cZ}{{\cal Z}}
\nc{\tr}{{{\rm tr}\,}}

\newcommand{\AmS}{{\protect\the\textfont2
  A\kern-.1667em\lower.5ex\hbox{M}\kern-.125emS}}

\title{Lattice QED and Universality of the Axial Anomaly
\thanks{Talk presented by H.J. Rothe at Lattice 99}}

\author{T.~Reisz$^{ab}$\thanks{Supported by a Heisenberg Fellowship} and 
H.~J.~Rothe$^b$
         \\ $^a$SPHT CEA-Saclay, Gif-sur Yvette Cedex, France
         \\$^b$Institut
        f\"ur Theoretische Physik, Universit\"at Heidelberg, Germany}
       
\begin{document}


\begin{abstract}
We give a perturbative proof that U(1) lattice gauge theories generate the 
axial anomaly in the continuum limit under very general conditions on the 
lattice Dirac operator. These conditions are locality, gauge covariance 
and the absense of species doubling. They hold for Wilson fermions 
as well as for realizations of the Dirac operator that 
satisfy the Ginsparg-Wilson 
relation. The proof is based on the lattice power counting theorem. 
The results generalize to non-abelian gauge theories. 
\end{abstract}

\maketitle

%
%

\section{Introduction}
In the continuum formulation of QED (or QCD) it is well known that all
known gauge
invariant regularization
schemes yield the ABJ axial anomaly. A non-perturbative regularization is
provided
by the lattice. Any candidate for a lattice discretization of the QED 
action should
reproduce
the axial anomaly in the continuum limit. A naive symmetric discretization of the continuum Dirac operator fails to
generate the anomaly, but is also plagued with the fermion doubling
problem. The
doubling problem can be eliminated by introducing an explicit,
naively irrelevant, chiral symmetry breaking Wilson term in the action.
This circumvents
the Nielsen Ninomyia theorem \cite{Nielsen} and leads to the
correct anomaly in the continuum limit \cite{Karsten,Rothe}.

The question arises whether this is a particular property of the
action chosen, or whether the anomaly will be correctly reproduced by any
lattice
regularization of the action satisfying some very general conditions. In
fact, as we
will show, any lattice action which is i) gauge invariant, ii)
possesses the correct continuum limit, iii) is local in some generalized
sense,  and iv) is free of doublers, will necessarily reproduce the axial
anomaly in the continuum limit. 
An important ingredient which will be needed for the proof is the general form
of the lattice axial vector Ward identity. Its precise
structure, which will depend on the particular discretization of the action
chosen,
need not be known, but only very general properties thereof. Our general
proof will
include the case of Ginsparg-Wilson fermions.

\section{Ward Identity}

Consider the following general form for the fermionic contribution to the lattice action
\be\label{action}
S_F[U,\psi,\bar{\psi}]  = \sum_{x,y} \bar\psi(x)(D_U(x,y)+m)\psi(y)\ .
\ee
Here $U$ stands for the collection of link variables, $D_U$ is the  
Dirac operator, and $\sum_x \equiv \sum_n a^4$, where $n$ label the lattice sites.  $D_U$ can be decomposed into a part
anticommuting with $\gamma_5$, which in the continuum limit takes the
form $\gamma\cdot\partial$, and a term commuting with
$\gamma_5$, which vanishes in the continuum limit.
Hence the general structure of the lattice axial vector Ward identity in an external gauge field will be of the form
\be\label{Wardidentity}
\langle\partial^*_\mu j^5_\mu(x)\rangle_U = 2m \; \langle j^5(x)\rangle_U+
\langle\Delta (x)\rangle_U 
\ee
where $\partial^*_\mu (\partial_\mu)$ stands for the dimensioned left (right) lattice derivative. $j^5(x)={\bar\psi}(x)\gamma_5\psi(x)$, and $j^5_\mu(x)$ has the correct continuum limit. $\Delta(x)$ is an "irrelevant" operator which vanishes in the naive continuum limit. For Wilson fermions $j^5_\mu$ and $\Delta$ have 
been given in \cite{Karsten}. Furthermore, it was shown in 
\cite{Karsten,Rothe}, that the anomaly is generated in the continuum limit by the
 naively "irrelevant" contribution $<\Delta(x)>$ .
For Ginsparg-Wilson fermions the Dirac operator in (\ref{action}) satisfies the
Ginsparg Wilson relation \cite{Ginsparg}
\[
\gamma_5 D + D\gamma_5 = a D\gamma_5 D \ , 
\]
where $a$ is the lattice spacing.  
The action possesses an exact global axial
symmetry \cite{Luescher1} for $m=0$ under the transformations 
\be\label{axialtr1}
\delta\psi(x) = i\omega\gamma_5[(1-{a\over 2}D)\psi](x)
\ee
\be\label{axialtr2}
\delta{\bar\psi}(x) = i\omega[{\bar\psi}(1-{a\over 2}D)](x)\gamma_5 \ .
\ee
According to the lattice Poincare Lemma \cite{Luescher2} there exists an axial vector
current associated
with this symmetry. Under a local transformation with $\omega$
replaced by $\omega(x)$,
the variation of the contribution to the fermionic action involving the
Dirac operator in (\ref{action}) can be written in the form
$\delta S_{D} = -i\sum_x \omega(x)\partial^*_\mu j^5_\mu(x)$. Taking into 
account the variation of the fermionic measure one then finds that
the axial Ward identity is of the
form (\ref{Wardidentity}), where the irrelevant $\Delta$ operator is given
by
\[
\Delta_{GW}(x) = a(\bar\psi D)(x)\gamma_5(D\psi)(x)\ . 
\]
Consider now the Ward identity (\ref{Wardidentity}) in momentum space. Define $<j^5_\mu(q)>_U$ by
\[
<j^5_\mu(x)>_U = \int^{{\pi\over a}}_{-{\pi\over
a}}{d^4q\over(2\pi)^4}<j^5_\mu(q)>_U
e^{i(q\cdot x+a\hat\mu/2)}
\]
Expanding $U_\mu(x) =
exp[igaA_\mu(x)]$ in the potentials, $<j^5_\mu(q)>_U$
will have the form
\begin{eqnarray}
\nonumber
&<j^5_\mu(q)>_U=\sum_n {1\over n!}\sum_{\{\nu_i\}}
\int_{\{k_i\}}\delta(\sum_ik_i-q)\\
\nonumber
&\times\Gamma^{5\mu}_{\nu_1\cdot\cdot\cdot\nu_n}(k_1,\cdot\cdot\cdot,k_n)
 A_{\nu_1}(k_1)\cdot\cdot\cdot A_{\nu_n}(k_n)\ ,
\end{eqnarray}
with analogous expressions for $<j^5(q)>_U$ and  $<\Delta (q)>_U$.
The Ward identity in momentum space then reads
\be
{\tilde q}_\mu\Gamma^{5\mu}_{\{\nu_i\}}(\lbrace k_i\rbrace)
= 2m\Gamma^{5}_{\{\nu_i\}}(\lbrace k_i\rbrace)
+ \Gamma^{(\Delta)}_{\{\nu_i\}}(\lbrace k_i\rbrace)\ , 
\ee
where ${\tilde q}_\mu = \frac{2}{a}\sin\frac{q_\mu a}{2}$.
\section{\bf Theorem}
We now state the central theorem of this paper and then give the proof:

Any lattice discretization of the QED action with the following properties: 
a) S has the correct continuum limit; b) S is gauge invariant; c) The Dirac Operator is local and d) The
free Dirac operator $D^{(0)}$(p) is free of doublers,
reproduces the axial anomaly in the continuum limit.

As we shall see below, the explicit form of
the axial vector current and of the irrelevant operator $\Delta(x)$ is not
required. 
\smallskip\noindent
Let us consider a) to c) in turn:

\noindent
 1) {\it S has the correct continuum limit}:
This ensures that $j^5_\mu$ posesses the correct naive continuum limit.

\noindent
2) {\it Gauge invariance}: Gauge invariance tells us that if ${\cal O}(\psi,{\bar\psi},A)$ is a gauge
invariant operator, then its external field expectation value satisfies
\[
<{\cal O}(\psi,{\bar\psi},A^\omega)>_{A^\omega} = <{\cal O}(\psi,{\bar\psi},A)>_A \ ,
\]\ 
where $A^\omega_\mu(x) = A_\mu(x) +{\partial}_\mu \omega(x)$.
It follows that
\[
\sum_\nu {\partial}^*_\nu {\partial\over \partial A_\nu(z)}
<{\cal O}(\psi,\bar\psi,A)>_A = 0\ ,
\]
or in momentum space
\be\label{gaugeinv1}
\sum_{\nu_i}({\tilde k}_i)_{\nu_i}
\Gamma^{(\cal O)}_{\nu_1\cdot\cdot\cdot\nu_n}(k_1,\cdot\cdot\cdot,k_n) =0\ .
\ee
 \noindent
3) {\it Locality of the Dirac Operator}:
The Dirac operator $D_U(x,y)$ can be formally expanded in a power series in the
gauge potentials:
\begin{eqnarray}
\nonumber
D_U(x,y)&=\sum_{n,\mu_i,z_i} {1\over n!}D^{(n)}_{\mu_1\cdot\cdot
\cdot\mu_n}(x,y|z_1\cdot\cdot\cdot z_n)\\
\nonumber
&\times A_{\mu_1}(z_1)\cdot\cdot\cdot
A_{\mu_n}(z_n)\ .
\end{eqnarray}
The coefficient functions $D^{(n)}_{\mu_1\cdot\cdot
\cdot\mu_n}$ are assumed to vanish exponentially fast with the separation
between any
pair of lattice sites with a decay constant of the order of the lattice 
spacing (this also holds for Ginsparg Wilson fermions for
sufficiently smooth
gauge field configurations \cite{Luescher3}. As a consequence
$\Gamma^{5\mu}_{\nu_1\cdot\cdot\cdot\nu_n}(\lbrace k_i\rbrace) \ ,\ \
\Gamma^{5}_{\nu_1\cdot\cdot\cdot\nu_n}(\lbrace k_i\rbrace)$, 
and $\Gamma^{(\Delta)}_{\nu_1\cdot\cdot\cdot\nu_n}(\lbrace k_i\rbrace)$
can be Taylor expanded around zero momenta. It then follows from (\ref{gaugeinv1}) 
that
\be\label{gaugeinv2}
{\cal T}_1 \Gamma_{\nu_1,\cdot\cdot\cdot,\nu_n}(k_1,\cdot\cdot\cdot,k_n)
= 0\ ,
\ee
where ${\cal T}_n$ denote the Taylor expansion up to order $n$ in the
momenta

\noindent
 4) {\it The free propagator} $D^{-1}_{(0)}(p)$ {\it is free of doublers}.
As a consequence the lattice power counting rules \cite{Reisz1} and the Reisz
theorem \cite{Reisz2} applies. This theorem allows one to take the naive 
continuum limit of a lattice integral, if all the lattice degree of divergences 
of the integrand, including the measure, are negative.

\noindent
{\it Proof of the Theorem}

In the following we only consider the axial vector Ward identity for two external
photons, since it is the only one which is anomalous. It has the form
\be\label{Wardidentity2}
{\tilde q}_\mu\Gamma^{5\mu}_{\nu_1\nu_2}(\{k_i\})
= 2m\Gamma^{5}_{\nu_1\nu_2}(\{k_i\})
+ \Gamma^{(\Delta)}_{\nu_1\nu_2}(\{k_i\})\ ,
\ee
where $i=1,2$. By power counting $degr \Gamma^{5\mu} = degr \Gamma^{5} = 1$,
while the irrelevant contribution has degr
$\Gamma^{(\Delta)} \le 2$. Here $degr F$ denotes the lattice divergence degree of $F$ \cite{Reisz1}.

We first show that all the vertex functions possess a continuum limit.
Consider e.g. $\Gamma^{5\mu}$, and write it in the form
\be\label{Taylor}
$$\Gamma^{5\mu}_{\nu_1\nu_2}=(1-{\cal{T}}_1)\Gamma^{5\mu}_{\nu_1\nu_2}
+{\cal{T}}_1\Gamma^{5\mu}_{\nu_1\nu_2}
\ee
The first term on the RHS has an $degr < 0$. Hence making use of the Reisz
theorem,
its continuum limit is given by the
Taylor subracted continuum one loop integral with a $\gamma_\mu\gamma_5$
insertion.
This is just the usual continuum triangle graph, Taylor subtracted to first
order in the momenta. The second term on the RHS of (\ref{Taylor}) vanishes
because of gauge invariance (\ref{gaugeinv2}). The $a\to 0$ limit of
$\Gamma^{5}_{\nu_1\nu_2}$ can be calculated
in a similar way. Hence we immediately arrive at the result that the
continuum limit
of $\Gamma^{(\Delta)}_{\nu_1\nu_2}$ in (\ref{Wardidentity2}) is universal, and given in terms of convergent continuum Feynman integrals. This limit can be 
easily computed. To this effect  
decompose $\Gamma^{(\Delta)}_{\nu_1\nu_2}$ as follows:
\[
\Gamma^{(\Delta)}_{\nu_1\nu_2} =
(1-{\cal{T}}_2)\Gamma^{(\Delta)}_{\nu_1\nu_2} +
{\cal{T}}_1\Gamma^{(\Delta)}_{\nu_1\nu_2}+
({\cal{T}}_2-{\cal{T}}_1)\Gamma^{(\Delta)}_{\nu_1\nu_2}
\]
where ${\cal{T}}_2$ denotes Taylor subtraction to second order in the momenta.
Because of gauge invariance (\ref{gaugeinv2}) the second term on the RHS vanishes. According
to the Reisz theorem \cite{Reisz2} the first term vanishes
for $a\to 0$ since degr[$\Gamma^{(\Delta)}]\le 2$, and
$\Gamma^{(\Delta)}$
vanishes in the naive continuum limit. Hence we conclude that
\[
\lim_{a\to 0}\Gamma^{(\Delta)}_{\nu_1\nu_2}(k_1,k_2) =
\lim_{a\to
0}({\cal{T}}_2-{\cal{T}}_1)\Gamma^{(\Delta)}_{\nu_1\nu_2}(k_1,k_2)
\]
We now make use of the Ward identity (\ref{Wardidentity2}) to obtain a simple expression for
the RHS. Thus
applying the ${\cal{T}}_2-{\cal{T}}_1$ operation to both sides of 
(\ref{Wardidentity2}),
making use of
gauge invariance (\ref{gaugeinv2}) and of the Reisz theorem, one readily finds that
\[
\lim_{a\to 0}\Gamma^{(\Delta)}_{\nu_1\nu_2} =
 2m\int{d^4\ell\over(2\pi)^4}\ ({\cal T}_1-{\cal T}_2)
I^5_{\nu_1\nu_2}(\ell,k_1,k_2)
\]
where $I^5_{\nu_1\nu_2}(\ell,k_1,k_2)$ is the integrand of the
continuum Feynman integral
for the triangle graph with a $\gamma_5$ insertion.
Hence the anomalous contribution reads:
\begin{eqnarray}
\nonumber
  {\cal A}_{\mu\nu}=-16e^2\epsilon_{\mu\nu\sigma\rho}k_{1\sigma}k_{2\rho}
\int {d^4\ell\over(2\pi)^4}\ {m^2\over{(\ell^2+m^2)^3}} && \\
\nonumber
 \qquad =-\frac{1}{2\pi^2}\sum_{\rho,\sigma}\epsilon_{\mu\nu\rho\sigma}
k_{1\rho}k_{2\sigma}\qquad\qquad\qquad\qquad &&
\end{eqnarray} 
We have thus shown that, irrespective of the detailed structure of the lattice
current and of the irrelevant operator $\Delta$ in the Ward identity 
(\ref{Wardidentity2}),
the correct anomaly is generated
by the irrelevant $\Delta$-contribution in the continuum limit. This limit
is universal
and given by the second order Taylor term in the expansion of the continuum
triangle graph with a
$\gamma_5$ vertex insertion. This expression has precisely the same form as
that obtained
in the continuum formulation using the Pauli-Villars regularization scheme,
except that here
the Pauli Villars mass is replaced by the fermion mass. The
integral is however independent of the mass. Furthermore we have seen that all 
terms in the Ward identity (\ref{Wardidentity2}) posssess a continuum limit. The limit of the first two terms is however {\it not} given by the naive limit 
of the triangle graphs, but actually by their BPHZ subtracted form.

\end{document}